\documentclass[%
 reprint,
superscriptaddress,
 amsmath,amssymb,
 aps,
 prd,
]{revtex4-2}

\usepackage{graphicx}
\usepackage{dcolumn}
\usepackage{bm}
\usepackage{xcolor}
\usepackage{orcidlink}

\DeclareRobustCommand{\VAN}[3]{#2}
\let\VANthebibliography\thebibliography
\def\thebibliography{\DeclareRobustCommand{\VAN}[3]{##3}\VANthebibliography}

\newcommand{\msun}{{\,\rm M_\odot}}

\newcommand{\kms}{\,{\rm km}\,{\rm s}^{-1}}

\newcommand{\cpm}{\,{\rm cm}^2\,{\rm g}^{-1}}

\def\jcap{J. Cosmol.  Astropart. Phys.}

\def\apj{ApJ}

\def\apjl{ApJ}
\def\mnras{MNRAS}

\def\physrep{Phys. Rep.}

\def\apjs{ApJS}
\def\prd{Phys. Rev. D}

\begin{document}

\preprint{APS/123-QED}

\title{Gravothermal Catastrophe in Resonant Self-interacting Dark Matter Models}

\author{Vinh Tran\,\orcidlink{0009-0003-6068-6921}}
 \affiliation{Department of Physics and Kavli Institute for Astrophysics and Space Research, Massachusetts Institute of Technology, Cambridge, MA 02139, USA}
 \email{vinhtran@mit.edu}
 
\author{Daniel Gilman\,\orcidlink{0000-0002-5116-7287}}
\affiliation{Department of Astronomy \& Astrophysics, University of Chicago, Chicago, IL 60637, USA}
\affiliation{Brinson Prize Fellow}

\author{Mark Vogelsberger\,\orcidlink{0000-0001-8593-7692}}
\affiliation{Department of Physics and Kavli Institute for Astrophysics and Space Research, Massachusetts Institute of Technology, Cambridge, MA 02139, USA}

\author{Xuejian Shen\,\orcidlink{0000-0002-6196-823X}}
\affiliation{Department of Physics and Kavli Institute for Astrophysics and Space Research, Massachusetts Institute of Technology, Cambridge, MA 02139, USA}

\author{Stephanie O'Neil\,\orcidlink{0000-0002-7968-2088}}
\affiliation{Department of Physics and Kavli Institute for Astrophysics and Space Research, Massachusetts Institute of Technology, Cambridge, MA 02139, USA}

\author{Xinyue Zhang\,\orcidlink{0000-0002-9502-5600}}
\affiliation{Department of Industrial Engineering and Operations Research, Columbia University, New York, NY 10027, USA}

\date{\today}

\begin{abstract}
    We investigate a self-interacting dark matter (SIDM) model featuring a velocity-dependent cross section with an order-of-magnitude resonant enhancement of the cross section at $\sim 16\kms$. To understand the implications for the structure of dark matter halos, we perform N-body simulations of isolated dark matter halos of mass $\sim 10^{8}\msun$, a halo mass selected to have a maximum response to the resonance. We track the core formation and the gravothermal collapse phases of the dark matter halo in this model and compare the halo evolving with the resonant cross section with halos evolving with velocity-independent cross sections. We show that dark matter halo evolution with the resonant cross section exhibits a deviation from universality that characterizes halo evolution with velocity-independent cross sections. The halo evolving under the influence of the resonance reaches a lower minimum central density during core formation. It subsequently takes about $20\%$ longer to reach its initial central density during the collapse phase. These results motivate a more detailed exploration of halo evolution in models with pronounced resonances. 
\end{abstract}

\maketitle


\section{Introduction}
\label{sec:intro}

To date, dark matter (DM) remains a purely astrophysical phenomenon~(see \cite{Bertone2005} for a review), with the only positive evidence for its existence coming from cosmological observations~\citep{Buckley++18}. In particular, some of the strongest bounds on the nature of DM from astrophysics derive from inferences of the properties of DM halos, the concentrations of DM that host galaxy clusters and galaxies in their centers. In many classes of DM theory, the details of the particle physics model alter the abundance and internal structure of halos, leading to constraints on DM particle physics from inferences of the abundance and internal structure of halos \citep{DrlicaWagner++22,Bechtol++22}.

The concordance cosmological model includes cold dark matter (CDM) which, by definition, behaves as a collisionless fluid. In the last two decades, self-interacting dark matter (SIDM) has emerged as a viable alternative to CDM \cite[e.g.][]{Carlson1992, DeLaix1995, Firmani2000, Spergel2000, Tulin13, Vogelsberger2012}. SIDM refers to a class of theory in which the DM has a non-zero self-interaction cross section~(see \cite{Adhikari+22} for a review). The exchange of energy and momentum through scattering in SIDM causes a temporal evolution in the internal structure of halos, whereas CDM predicts a static internal structure well-approximated by a Navarro-Frenk-White profile \citep[][hereafter NFW profile]{NFW1997} for halos at all times. The evolution of halo internal structure begins with a period of core formation \cite[e.g.][]{Vogelsberger2012, Rocha2013, Zavala2013, Elbert2015}, during which time heat flows into the center of the halo. Following core formation, the direction of heat flow eventually reverses, causing a runaway contraction of the density profile referred to as `gravothermal catastrophe', or core collapse. This phenomenon was initially studied in the context of dense star clusters \citep{LyndenBell68}. Several decades later, \citet{Balberg2002} (see also \citet{Koda2011}) pointed out that DM self-interactions provide a more efficient mechanism to transfer heat throughout halo profiles than purely gravitational scattering. Core collapse represents the final state of isolated self-interacting DM halos \cite[e.g.][]{Yang+2021,Yang+2022,Nadler+23,Outmezguine+2023}. Inside galactic halos, various processes such as tidal stripping and scattering between subhalo and host halo particles can delay or accelerate the onset of collapse in subhalos \citep{Nishikawa+20,Nadler+20,Sameie2020,Zeng+22,Slone+23,Zeng+23}. More complicated core collapse behaviors are found under the influence of baryons~\citep[e.g.][]{Feng2021} or with dissipative DM self-interactions~\citep[e.g.][]{Essig2019,Vogelsberger2019,Shen2021,ONeil2023}.

Core formation and collapse in SIDM theories result in a variety of observable consequences for the properties of dwarf galaxies~\citep[e.g.][]{Kaplinghat+16,Tulin2018,Silverman+23,Nadler+23,Shah+24}. In addition, cored and collapsed halos produce distinct gravitational lenses signatures that enable studies of SIDM from observations of strongly-lensed quasars \citep{Gilman+21,Gilman+23}, as well as galaxy-galaxy strong lenses \citep{Minor+21}. The most definitive existing constraints on SIDM theories come from observations of galaxy clusters, which place stringent upper limits on the self-interaction cross section $\sigma < 0.1 \rm{cm^2} \ \rm{g^{-1}}$~\cite{Peter+2013,Harvey2015,Sagunski+2021}, effectively requiring that the DM behave as a collisionless fluid on halo mass scales $\sim 10^{14} M_{\odot}$. The total mass of a collapsed halo sets the velocity scale for the scattering process, as the typical speed of a DM particle in a virialized structure increases with the mass of the halo. Thus, the constraints on the self-interaction cross section from galaxy clusters place stringent upper limits on the self-interaction cross section at relative velocities $v \sim 1000 \ \rm{km} \ \rm{s^{-1}}$. For a velocity-independent scattering cross section, this upper limit would rule an SIDM model capable of altering the observable properties of DM halos on galactic and sub-galactic scales. However, if the SIDM cross section has a velocity dependence that increases the cross section at low relative velocities, heat transfer in low-mass halos can become significantly enhanced, and core formation and collapse occur within the age of the Universe~\citep[e.g.][]{Correa2021,Turner2021}. Velocity-dependent cross sections that become amplified at low speeds are a common feature of scattering mediated by light force carriers with Yukawa interaction potentials (for a detailed review of this topic, see ~\citet{Tulin13}). 

The current modeling of core collapse assigns a DM halo a single effective cross section as a function of the velocity-dependent cross section and the halo's central velocity dispersion. From this effective cross section, one obtains a characteristic timescale for which the evolution of SIDM halos is self-similar, or universal \citep{Yang+2022,Outmezguine+2023,Yang+23}. This approach enables a predictive framework for modeling core collapse, enabling the determination of the time of core collapse as a multiple of the collapse timescale given the properties of the SIDM cross section and structural parameters of a halo \citep[e.g.][]{Gilman+21}. One potential complication for this relatively simple picture of halo evolution in SIDM is that for a DM halo of a fixed total mass, the typical relative velocity between particles changes as a function of radius, and thus the scattering inside a given halo occurs over a range of velocities. While some numerical simulations have verified the universality of halo evolution for some velocity-dependent cross sections \citep{Yang+2022}, in this work we consider a class of SIDM models with velocity-dependent cross sections that change significantly across the range of relative velocities typical of particles inside a halo. 

We consider a class of SIDM models in which a resonance in the cross section causes an order-of-magnitude enhancement to the scattering cross section near a particle relative velocity. Resonances refer to an enhancement or suppression of the cross section that results from the formation of bound states between DM particles and attractive potential. Resonances can manifest as single pronounced peaks, multiple peaks, or as suppression or enhancement of the cross section strength across a range of speeds, depending on the details of the particle physics model \citep{Tulin13,Chu++20,Gilman+23}. For this class of models, the cross section amplitude can significantly change with radius inside a halo, complicating the relatively simple physical picture in which a halo evolves under the influence of a single effective cross section. This class of models has received relatively limited study. Ref. \cite{Kamada+24} examined the evolution of a halo analytically using the gravothermal fluid formalism, while Ref. \cite{Zeng+23} studied a resonant model in a numerical simulation, but did not distinguish its structural evolution from models with simpler velocity-dependence.  

To gain a detailed physical understanding of the evolution of self-interacting DM halos with pronounced resonances in the cross section, we conduct idealized N-body simulations of isolated halos with a resonant cross section. We pick a halo mass expected to maximize the effect of the resonance on the halo evolution, and compare the evolution of the halo with the resonant cross section with the evolution of halos evolving under the influence of velocity-independent cross sections. We examine whether existing treatments of SIDM halo evolution and universality hold for this class of models. 

This paper is organized as follows: Section \ref{sec:particle_physics} gives an overview of the particle physics model and the calculation of the resonant cross section considered in this work. Section \ref{sec:analytical} discusses the calculation of two effective scattering cross sections we will consider when interpreting the evolution of the halo with a resonant cross section. In Section \ref{sec:simulation}, we review the details of our N-body simulations. Section \ref{sec:results} presents the results of the simulations. We summarize our findings and make concluding remarks in Section \ref{sec:conclusions}. 

\section{Particle physics model}
\label{sec:particle_physics}
We consider a SIDM cross section that results from interactions between DM particles of mass $m_{\chi}$ through a force carrier of mass $m_{\phi}$. We consider the case of an attractive Yukawa potential $\left(c = \hbar = 1\right)$
\begin{equation}
\label{eqn:yukawapot}
V\left(r\right) = \frac{-\alpha}{r}\exp{\left(-r m_{\phi}\right)}
\end{equation}
where $r$ is the distance between DM particles, and $\alpha$ determines the strength of the potential. While no exact closed-form solutions exist for the differential scattering cross section corresponding to $V\left(r\right)$, in certain regimes of parameter space various approximations give analytic and closed-form solutions that match the exact cross section to high precision. For example, for weak potentials $\left(\alpha m_{\chi} / m_{\phi} << 1\right)$, one obtains an approximate analytic expression for the differential scattering cross section using the Born approximation 
\begin{equation}
\frac{{\rm d} \sigma}{{\rm d} \Omega} = \frac{\sigma_0}{\left(w^2 + v^2 \sin^2 \left(\theta / 2\right)\right)^2}
\end{equation}
where $v$, $\theta$, and $\Omega$ represent relative velocity, scattering angle, and the solid angle, respectively. The parameters $\sigma_0$ and $w$ are related to the particle physics model by $\sigma_0 \equiv \alpha^2 m_{\chi}^2 / m_{\phi}^4$ and $w \equiv m_{\phi} / m_{\chi}$. 

Outside the range of validity of the Born approximation, one must compute the differential scattering cross section using partial wave analysis. In this case, the differential scattering cross section is given by 
\begin{equation}
\label{eqn:partialwavesum}
\frac{{\rm d} \sigma}{{\rm d} \Omega} = \frac{1}{k^2} \Big | \sum_{\ell = 0}^{\infty} \left(2 \ell + 1\right)e^{i \delta_l} \sin \delta_l P_{\ell}\left(\cos\theta\right)\Big |^2
\end{equation}
where $\delta_{\ell}$ represents the phase shifts of partial waves labeled by their angular momentum quantum number $\ell$, $k = m_{\chi} v / 2$ is the momentum of the DM particle, and $P_{\ell}\left(x\right)$ are Legendre polynomials of order $\ell$. 

It has become common practice to average over the angular dependence of the differential scattering cross section when predicting the internal structure of DM halos in SIDM. This simplification can yield accurate predictions for the evolution of halo profiles provided one chooses a suitable proxy for ${\rm d} \sigma / {\rm d} \Omega$. As shown by \citet{Yang+2022}, the viscosity transfer cross section $\sigma_V = \int \frac{{\rm d} \sigma}{{\rm d} \Omega} \sin^2\theta {\rm d} \Omega$, for which $\sigma_V$ is given by \citep{Colquhoun+21}
\begin{equation}
\label{eqn:viscositycross}
\sigma_V = \frac{4 \pi}{k^2} \sum_{\ell = 0}^{\ell_{\rm{max}}} \frac{\left(\ell + 1\right)\left(\ell +2\right)}{2 \ell + 3} \sin^2\left(\delta_{\ell +2} - \delta_{\ell}\right),
\end{equation}
allows one to predict the structural evolution of a halo without explicitly accounting for the angular dependence of the scattering cross section in N-body simulations. This represents a considerable simplification for the numerical scattering implementation, and for obtaining analytical predictions for the structural evolution of the halo. We truncate the summation at a value of $\ell_{\rm{max}}$ for which $\sigma_V$ converges to within $1 \%$


For an attractive potential given by Equation \ref{eqn:yukawapot} in the quantum regime $\left(k / m_{\phi} < 1\right)$, the formation of bound states with the potential gives rise to resonances in the cross section. A resonance refers to the enhancement or suppression of the scattering cross section across a range of relative velocities. The velocities where resonances in the cross section occur depend on the values of $\alpha$ and $m_{\phi} / m_{\chi}$ for the potential in Equation \ref{eqn:yukawapot}. Due to the computational cost of solving the full scattering problem and then running N-body simulations that implement the resulting cross sections, SIDM models with significant resonances have not been extensively studied in the literature. As a result, we lack a detailed understanding of structure formation in this class of SIDM theory, and it is unclear whether analytical treatments of halo evolution in SIDM calibrated for other velocity-dependent cross sections \citep[e.g.][]{Yang+2022,Yang+23} apply to this class of models.

To evaluate the cross section in the resonant regime, we solve the Schr{\"o}dinger equation to compute the phase shifts $\delta_{\rm{\ell}}$, and use Equation \ref{eqn:viscositycross} to compute the cross section. To calculate $\delta_{\ell}$ we use an auxiliary function for the Schr{\"o}dinger equation \citep{Chu++20}
\begin{widetext}
\begin{equation}
\dfrac{\partial \delta_{\ell}\left(r\right)}{\partial r} = -k\, m_{\chi}\, r^2\, V\left(r\right)\, \left[\cos\left(\delta_{\ell}\left(r\right)\right)\, j_{\ell}\left(kr\right) - \sin\left(\delta_{\ell}\left(r\right)\right)\, n_{\ell}\left(kr\right)\right]^2,
\end{equation}
\end{widetext}
where $j_{\ell}\left(x\right)$ and $n_{\ell}\left(x\right)$ are spherical Bessel functions of the first and second kind of order $\ell$. We obtain a numerical solution for $\delta_{\ell}\left(r\right)$ using {\tt{Mathematica}} with the initial condition $\delta_{\ell}\left(0\right)=0$, and obtain the phase shift by taking the limit $\delta_{\ell} = \lim_{r\rightarrow \infty} \delta_{\ell}\left(r\right)$. 

For the analysis presented in this work, we consider the resonant cross section represented by the shaded region in the left-hand panel Figure \ref{fig:collision_df}. This cross section was also considered by Ref. \cite{Gilman+23} in a strong-lensing analysis of multiply-imaged quasars, and has $m_{\rm{\chi}} = 31.8 \ \rm{GeV}$, $m_{\rm{\phi}} = 5.7 \ \rm{MeV}$, and $\alpha = 1.6 \times 10^{-3}$. The amplitude of the cross section, as indicated by the axis labels on the right-hand side of the figure, reaches $\sim 100 \ \rm{cm^2} \ \rm{g^{-1}}$ at $v \sim 16 \ \rm{km} \ \rm{s^{-1}}$, a feature that arises primarily from constructive interference between the $\ell = 3$ and $\ell = 1$ partial waves. To explore the consequences of this model for structure formation, we run a N-body simulation of a DM halo evolving under the influence of this resonant cross section, and track its evolution from core formation until core collapse.

\section{Analytic predictions for the evolution of halo structure}
\label{sec:analytical}
\begin{figure*}
    \centering
        \includegraphics[width=1.0 \textwidth]{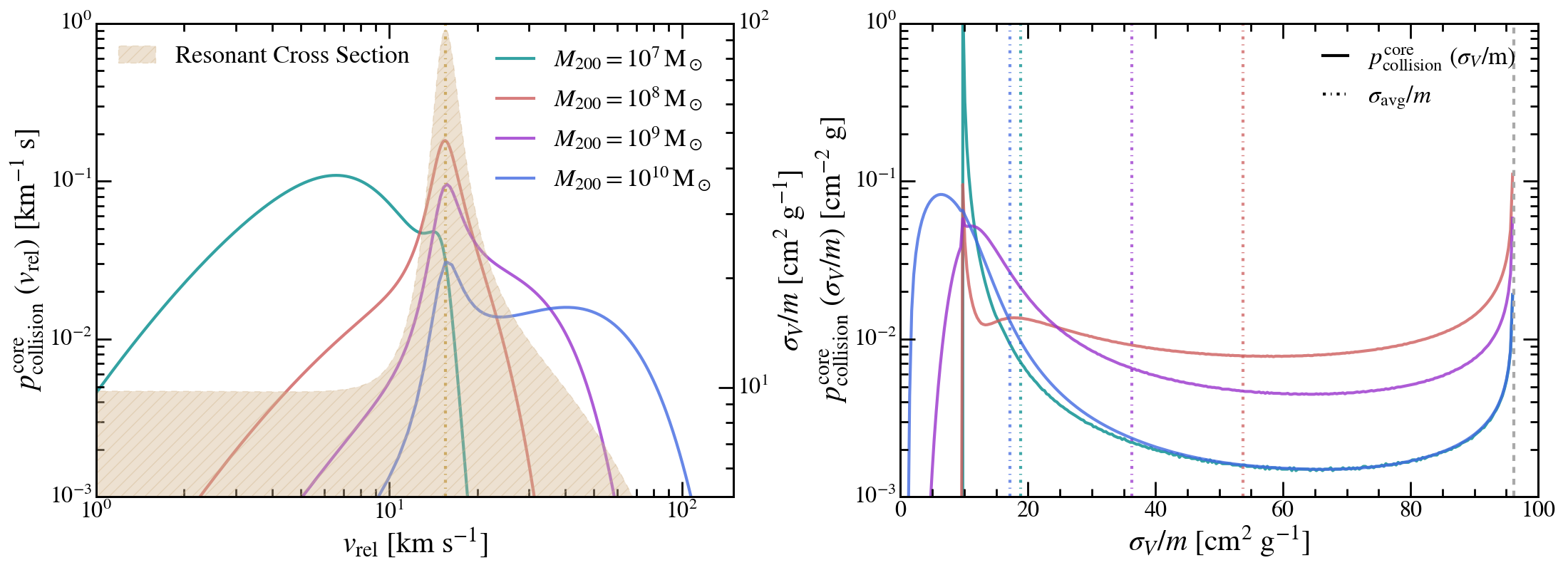}
    \caption{Particle collisions distribution as functions of relative velocity (left) and cross section per DM mass (right) for halos of masses $10^7$ (cyan), $10^8$ (red), $10^9$ (purple), and $10^{10}$ (blue) $\rm M_\odot$. The resonant cross section model is shown by the yellow profile in the left-hand panel, with the right-hand y-axis labels showing the amplitude of the cross section in $cm^2 \ \rm{g^{-1}}$. The effect of the resonant cross section can be observed clearly in the $10^8$ and $10^9\msun$ halos with strong peaks at the resonant relative velocity in the particle collision distribution. In the right figure, the grey vertical dashed line shows the upper limit of the cross section model. The vertical dash dotted lines indicate the collision rate effective cross section of the respective halo masses. The delta function-like features are due to the cross section of the plateau region at low relative velocity. From both figures, with the particle collision distribution and the effective cross sections, it is clear that the halo mass that maximizes the number of collisions happening around the resonant peak is around $10^8\msun$.}
    \label{fig:collision_df}
\end{figure*}

\begin{figure}
    \centering
        \includegraphics[width=0.49 \textwidth]{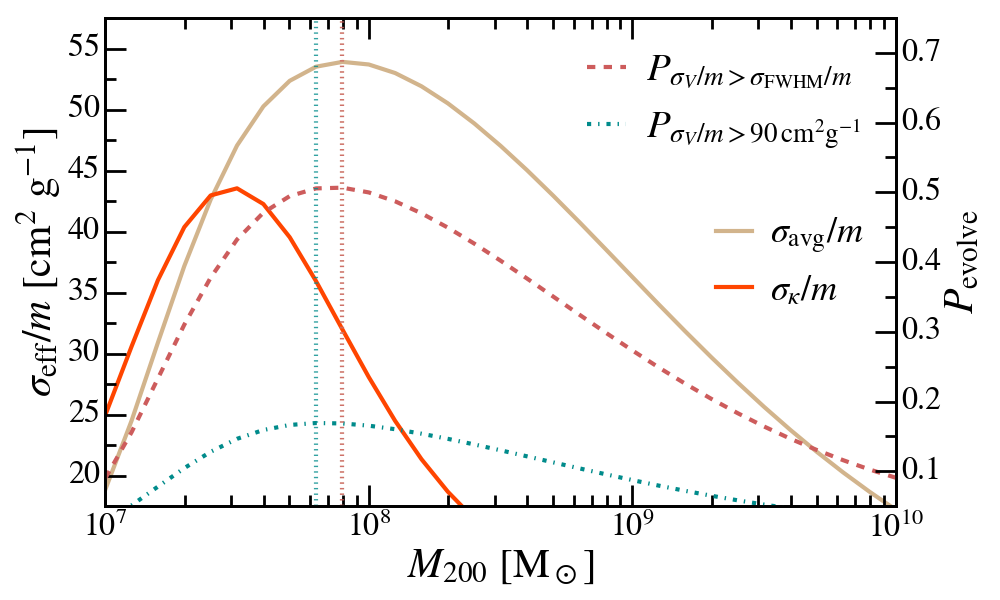}
    \caption{The left y-axis with the yellow and orange solid lines show the average and conductivity effective cross section (calculated from Equation \ref{eqn:average_cross_section} and \ref{eqn:conductivity_cross_section}, respectively) as functions of halo mass. The right y-axis and the two dashed lines show the portion of each halo evolving under a specific cross section range as a function of halo mass. The red line accounts for cross section within the FWHM of the resonant peak compared to the cross section of the plateau region at low relative velocity, while the blue line accounts for cross section near the resonant peak ($\sigma_V/m > 90\cpm$). These profiles peak at the halo of mass $10^{7.9}\msun$ and $10^{7.8}\msun$, respectively, although it must also be noted that the values for these two halo masses are very similar.}
    \label{fig:effective_cross_section}
\end{figure}

In the current picture of SIDM halo physics, the change in a halo density profile over time follows a self-similar or universal trajectory when the central density is scaled by $\rho_{\rm s}$, or a characteristic central density, and time is scaled by a characteristic timescale, $t_{\rm col}$, often referred to as the collapse timescale \citep{Outmezguine+2023,Yang+2022,Zhong+23}. This timescale is given by \cite{Yang2024}
\begin{equation}
    \label{eqn:collapse_time_scale}
    t_{\rm col}(C, \sigma_{\rm{eff}}/m) = \frac{150}{C} \frac{1}{\left(\sigma_{\rm eff}/m\right) \rho_{\rm s} r_{\rm s}} \frac{1}{\sqrt{4 \pi G \rho_{\rm s}}}.
\end{equation}
By convention, the numerical prefactor is expressed as $150 / C$ with a value set to match predictions from SIDM fluid models \cite[e.g.][]{Essig2019,Yang+23,Zhong+23} with the results from N-body simulations \citep[e.g.][]{Koda2011,Essig2019,Shen2021,Yang+2022}. In the current picture of SIDM halo evolution, $C$ is an $\mathcal{O}\left(1\right)$ constant, and all SIDM and thermodynamic considerations enter in the derivation of the effective cross section $\sigma_{\rm{eff}}$.

To test the hypothesis that SIDM halo evolution is universal when expressed in terms of $t_{\rm{col}}$, we perform N-body simulations of isolated halos with different cross section models, and compare their evolution using different definitions of $\sigma_{\rm{eff}}$. We evolve one halo with the resonant cross section shown in the left-hand panel of Figure \ref{fig:collision_df}. In addition, we simulate two other halos, the first evolving with a velocity-independent cross section with an amplitude $\sigma_{\rm{eff}}$ calculated by considering the scattering rate in a halo (see Section \ref{ssec:sigmav}), and the second halo with $\sigma_{\rm{eff}}$ calculated based on the heat conduction cross section $\langle \sigma v^5\rangle$ (see Section \ref{ssec:sigmav5}). If the universality of halo evolution holds for resonant models, the relation $\rho / \rho_s$ versus $t / t_{\rm{col}}\left(C,\sigma_{\rm{eff}}\right)$ should appear almost identical among the various simulations, modulo modest numerical systematic effects that affect the precise onset of core collapse \citep[e.g.][]{Mace+24}. 

\subsection{The collision rate effective cross section}
\label{ssec:sigmav}
In contrast to a velocity-dependent cross section, the resonant cross section shown on the left-hand side of Figure \ref{fig:collision_df}, exhibits a pronounced peak that increases the cross section amplitude by a factor of 10 at $\sim 16 \rm{km} \ \rm{s^{-1}}$. Therefore, we expect the most interesting structure formation consequences in halos that have the portion of scattering occurring near the peak of the resonance reach the maximum. To find the halo mass that maximizes scattering near the peak, we derive the particle collision distribution $p_{\rm collision}^{\rm core}\left(v_{\rm{rel}}\right)$, which quantifies the fraction of scattering events in a halo occurring in a range of relative velocities $\left[v_{\rm rel},v_{\rm rel} + {\rm d}v_{\rm rel}\right]$. From this distribution, we will also calculate $p_{\rm collision}^{\rm core}\left(\sigma_V / m\right)$, which represents the probability that a given particle in a halo sees a cross section amplitude $\sigma_V$. 

We assume an isotropic NFW DM density profile~\cite{NFW1997} 
\begin{equation}
    \label{eqn:nfw_inner_density}
    \rho\left(r\right) = \frac{\rho_{\rm s}}{\left(r/r_{\rm s}\right) \left(1 + \left(r/r_{\rm s}\right)\right)^2},
\end{equation}
where $\rho_{\rm s}$ is the scale density, and $r_{\rm s}$ is the scale radius. These parameters are calculated from the virial radius $r_{200}$, defined as the radius at which the average density of the halo is 200 times the critical density of the universe $\rho_{\rm crit}\,=\,3H^2/8\pi G$, and the concentration parameter $c$, with $c = r_{200} / r_{\rm s}$. Both of these parameters can be obtained with the virial mass $M_{200}$, the mass enclosed within $r_{200}$ - the former straightforwardly from its definition and the latter from the low-mass concentration-mass (c-M) relation at redshift $z=0$ following \cite{DiemerJoyce2019}. Thus, we only need the virial mass $M_{200}$ to parameterize our halos.

For dynamical equilibrium and the velocity distribution, we use Eddington's inversion formula~\cite{Binney1987}
\begin{equation}
    \label{eqn:eddington_inversion}
    \rho\left(r\right) = \int_0^{v_{\rm esp}} 4 \pi v^2 {\rm d}v\,f\left( \xi \right),
\end{equation}
where $v_{\rm esp}$ is the escape velocity at radius $r$, $\xi = \psi \left(r\right) - v^2/2$ is the binding energy per unit of mass, $\psi$ is the negative gravitational potential, and $f\left( \xi \right)$ is a phase-space distribution function. Here, we assume that the system is bounded and isotropic, thus $f$ only depends on $\xi$, and $f\left(<0\right) = 0$. The Eddington distribution function satisfying Equation \ref{eqn:eddington_inversion} takes the form
\begin{equation}
    \label{eqn:eddington_df}
    f \left( \xi \right) = \frac{1}{\sqrt{8} \pi^2} \int_{0}^{\xi} {\rm d}\psi \left[ \frac{1}{\sqrt{\xi-\psi}}\frac{{\rm d}^2\rho}{{\rm d}\psi^2} + \frac{1}{\sqrt{\xi}}\frac{{\rm d}\rho}{{\rm d}\psi} \Bigr\rvert_{\psi=0} \right].
\end{equation}
This can be evaluated from the analytical forms of $\rho\left(r\right)$ and $\psi\left(r\right)$ with $r$ as an intermediate variable. Integrating, we obtain the semi-analytical form of the Eddington distribution function. 


From the Eddington distribution function (Equation \ref{eqn:eddington_df}), we get the particle velocity distribution function at radius $r$ following
\begin{equation}
    \label{eqn:shell_velocity_df}
    p_{v}^{\rm shell}\left(v \mid r\right) = 4\pi v^2 \frac{f\left(\psi\left(r\right)-\frac{v^2}{2}\right)}{\rho\left(r\right)}.
\end{equation}
The particle relative velocity distribution function of particles within a thin shell at radius $r$ and width ${\rm d}r$ can then be retrieved using the equation
\begin{widetext}
\begin{equation}
    \label{eqn:shell_relative_velocity_df}
    p_{v_{\rm rel}}^{\rm shell}\left(v_{\rm rel} \mid r\right) = \int p_{v}^{\rm shell}\left(v_1 \mid r\right) {\rm d}v_1\,p_{v}^{\rm shell}\left(v_2 \mid r\right) {\rm d}v_2\,p_{\theta}\left(\theta\right)\,{\rm d}\theta\,\delta\left(v_{\rm rel} - \sqrt{v_1^2 + v_2^2 - 2v_1v_2\cos{\theta}}\right),
\end{equation}
\end{widetext}
with $v_1$, $v_2$ as the velocities of the particle pair, and $\theta$ as the relative angle between them. We assume the particle direction of movement as isotropic, which results in the distribution function of $\theta$ as follows
\begin{equation}
    \label{eqn:relative_angle_df}
    p_\theta\left(\theta\right) = \sin{\theta} / 2.
\end{equation}
To simplify calculations, we make use of the distribution of $\cos{\theta}$ rather than $\theta$. This distribution function takes the form of
\begin{equation}
    \label{eqn:cos_relative_angle_df}
    p_{\cos{\theta}}\left(\cos{\theta}\right) = 1 / 2
\end{equation}
and is a uniform distribution with the range of $\left[-1,1\right]$.

Simply integrating a complex triple integral such as Equation \ref{eqn:shell_relative_velocity_df} is unnecessarily complicated analytically and undesirably costly numerically. Therefore, we make use of a numerical trick involving the probability point functions (PPF) - the inverse function of the cumulative distribution function (CDF) - of $p_{v}^{\rm shell} \left(v \mid r\right)$ and $p_{\cos{\theta}}\left(\cos{\theta}\right)$. We create a 1500x1500x150 grid, and within each cell of the grid, we assign a $\left(v_1,v_2,\cos{\theta}\right)$ set of values which are decided by applying the appropriated PPF to a linearly-spaced array of the range $\left[0,1\right]$. What we do here is essentially similar to combining the three distribution functions, $p_{v}^{\rm shell}\left(v_1 \mid r\right)$, $p_{v}^{\rm shell}\left(v_2 \mid r\right)$, and $p_{\cos{\theta}}\left(\cos{\theta}\right)$, into a simultaneous distribution of all three values and discretize it. We then calculate the relative velocity, $v_{\rm rel}$, in each cell and bin the grid's data into a histogram. This functions similarly to integrating the distribution with the delta function, and by normalizing the histogram, we acquire the distribution function of the relative velocity inside a thin shell situated at radius $r$.

From the relative velocity distribution (Equation \ref{eqn:shell_relative_velocity_df}) we obtain the particle collisions distribution as a function of relative velocity $v_{\rm rel}$ using the following formula:
\begin{equation}
    \label{eqn:shell_collision_df}
    p_{\rm collision}^{\rm shell} \left(v_{\rm rel} \mid r\right) = C p_{v_{\rm rel}}^{\rm shell} \left(v_{\rm rel} \mid r\right) R \left(v_{\rm rel},r\right),
\end{equation}
where C is the normalizing factor; $R \left(v_{\rm rel},r\right)$ is the particle rate of collision, which follows
\begin{equation}
    \label{eqn:collision_rate}
    R \left(v_{\rm rel},r\right) = v_{\rm rel}\,\sigma_{V}\left(v_{\rm rel}\right) \rho\left(r\right),
\end{equation}
where $\sigma_{V}\left(v_{\rm rel}\right)$ is the cross section at the relative velocity $v_{\rm rel}$.

The particle collision distribution for the halo core can then be obtained by taking the weighted average of the thin shells' particle collision distributions with the shell's mass
\begin{equation}
    \label{eqn:core_collision_df}
    p_{\rm collision}^{\rm core} \left(v_{\rm rel}\right) = C^\prime \int_0^{3 r_{\rm s}} 4 \pi r^2 {\rm d}r \rho\left(r\right) p_{\rm collision}^{\rm shell} \left(v_{\rm rel} \mid r\right),
\end{equation}
where $C^\prime$ is another normalizing factor. Here we take the halo core to be the region within 3\,$r_{\rm s}$. Outside of this region, the number density is low enough that collision rarely happens. We also compute the collision distribution as a function of cross section $p_{\rm collision}^{\rm core} \left(\sigma_V\right)$, which one can interpret as the probability that a particle inside the halo sees a cross section amplitude $\sigma_V$. We compute this distribution by sampling from $p_{\rm collision}^{\rm core} \left(v_{\rm rel}\right)$, and evaluating $\sigma_V$ at the resulting relative velocities. The particle collision distribution as a function of relative velocity, $v_{\rm rel}$, and as a function of cross section per mass, $\sigma_V/m$, for halos of mass $10^7$, $10^8$, $10^9$, and $10^{10}\msun$ are shown in Figure \ref{fig:collision_df} with our resonant cross section model.

In Figure \ref{fig:collision_df}, we notice the distinguished peaks in the collision distribution at the resonant relative velocity for the higher mass halos and a fainter sign for the lower mass halo in the left figure. In the right figure, we observe a delta function-like feature at the cross section of the plateau region at low relative velocity for all halo masses, though the feature weakens as the mass increases and almost disappears as the halo mass reaches $10^{10}\msun$. At higher cross sections, the probability density increases with the halo mass reaching its maximum around the $10^8\msun$ halo. The probability density then decreases as the mass increases further. All of these features are expected from the features of our resonant cross section model.

From the core's particle collision distribution (Equation \ref{eqn:core_collision_df}), we acquire the average cross section seen by the halo following the formula
\begin{equation}
    \label{eqn:average_cross_section}
    \sigma_{\rm avg} = \int {\rm d}v_{\rm rel}\,p_{\rm collision}^{\rm core} \left(v_{\rm rel}\right) \sigma_V\left(v_{\rm rel}\right).
\end{equation}
Onward we called this cross section the collision rate effective cross section. 

\subsection{The heat conduction cross section}
\label{ssec:sigmav5}
Additionally, existing treatments of gravothermal evolution with certain classes of velocity-dependent cross section evolve in a self-similar way when computed with a heat conductivity motivated effective cross section $\sigma_\kappa$, assuming a Maxwell-Boltzmann distribution. We also consider this approach and inspect the effective cross section \cite{Yang+2022,Yang+23}
\begin{equation}
    \label{eqn:conductivity_cross_section}
    \sigma_{\kappa} = \frac{\langle \sigma_V v^5 \rangle}{\langle v^5 \rangle} = \frac{\int \sigma_V\left(v\right) v^5 p_v^{\rm MB}\left(v\right) {\rm d}v}{\int v^5 p_v^{\rm MB}\left(v\right) {\rm d}v}.
\end{equation}
Here, $p_v^{\rm MB}\left(v\right)$ is the Maxwell-Boltzmann distribution
\begin{equation}
    \label{eqn:Maxwell-Boltzmann_distribution}
    p_v^{\rm MB}\left(v\right) \propto v^2 \exp\left(-\frac{v^2}{4 v_0^2}\right),
\end{equation}
where $v_0 = 0.64\,v_{\rm max}$ with $v_{\rm max} = 1.65\sqrt{G \rho_{\rm s} r_{\rm s}^2}$ as the maximum circular velocity. The average and conductivity effective cross sections of halos with masses in the range of $10^7$ to $10^{10}\msun$ are displayed in Figure \ref{fig:effective_cross_section}. In the figure are also two lines indicating the portion of each halo that will evolve under cross sections within the full width at half maximum (FWHM) of the resonant peak (compared to the cross section of the plateau region at low relative velocity) $P_{\sigma_V/m > \sigma_{\rm FWHM}/m}$, and the portion of each halo that will evolve under cross sections near the resonant peak $P_{\sigma_V/m > 90\cpm}$. We observe that halo with mass of $10^{7.9}\msun$ have the largest effective cross section and the largest $P_{\sigma_V/m > \sigma_{\rm FWHM}/m}$, while halo with mass of $10^{7.8}\msun$ have the largest $P_{\sigma_V/m > 90\cpm}$. Onward, we will keep our focus on the evolution of the $10^{7.9}\msun$ halo, which has an effective cross sections of $\sigma_{\rm avg}/m = 53.91\cpm$ and $\sigma_\kappa/m = 31.98\cpm$. Other parameters of the halo are shown in Table \ref{tab:runs_config}.

\section{Simulations}
\label{sec:simulation}

\begin{figure}
    \centering
        \includegraphics[width=0.49 \textwidth]{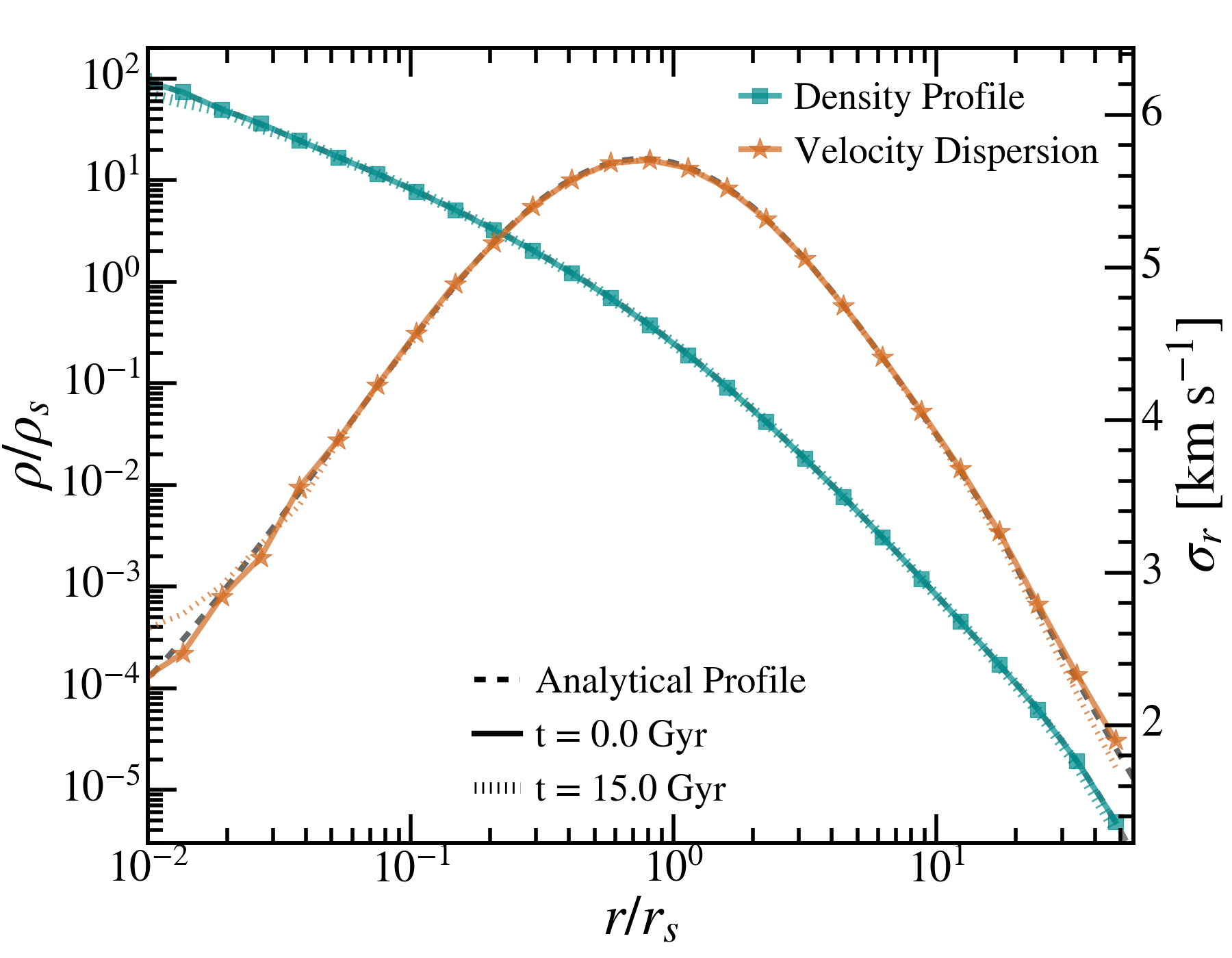}
    \caption{Density (blue, left) and velocity dispersion (orange, right) profiles of the $10^{7.9}\msun$ CDM halo. The black dashed lines show the halo analytical profiles following Equation \ref{eqn:nfw_inner_density}, \ref{eqn:nfw_outer_density}, and \ref{eqn:velocity_dispersion}. The colored solid lines and the colored dotted lines show the profiles at 0\,Gyr and 15\,Gyr, respectively. The formers are generated from the initial conditions sampling code and used as the initial conditions of the CDM simulation. The laters show signs of small fluctuations in the innermost region of the halo due to the statistical error arising from the limited number of particles within the region, as well as fluctuations in the outer edge of the halo during the halo's evolution to stability. There also exists a small core due to the effect of the softening length. However, the deviation from the analytical profiles resulting from this small core is of the same order of magnitude as the statistical fluctuation. The evolved 15\,Gyr CDM halo is then used as the initial conditions for the SIDM runs.}
    \label{fig:ics}
\end{figure}

\begin{table*}
    \centering
    \addtolength{\tabcolsep}{9pt}
    \def\arraystretch{1.2}
    \begin{tabular}{c c c c c c c c c c}
        \hline
        $\log M_{200}$ & $N_{200}$ & $m_{\rm DM}$ & $r_{200}$ & $c$ & $r_{\rm s}$ & $\log \rho_{\rm s}$ & $\epsilon$ & $\sigma \, m_{\rm \chi}^{-1}$ & $t_{\rm col}$\\ [0.25ex] 
        [${\rm M}_\odot$] &  & [${\rm M}_\odot$] & [$\rm kpc$] &  & [$\rm kpc$] & [${\rm M}_\odot\,{\rm kpc}^{-3}$ ] & [$\rm pc$] & [${\rm cm}^2\,{\rm g}^{-1}$] & [Gyr]\\ [1ex] 
        \hline\hline
        7.9 & $3 \times 10^7$ & 2.64 & 9.08 & 18.7 & 0.49 & 7.43 & 1.00 & Resonant & Vary \\
        \hline
        7.9 & $3 \times 10^7$ & 2.64 & 9.08 & 18.7 & 0.49 & 7.43 & 1.00 & $\sigma_{\rm{avg}}=53.91$ & 22.94 \\
        \hline
        7.9 & $3 \times 10^7$ & 2.64 & 9.08 & 18.7 & 0.49 & 7.43 & 1.00 & $\sigma_{\kappa}=31.98$ & 38.67 \\ [0.25ex] 
        \hline
    \end{tabular}
    \caption{Simulations configuration. (1) $M_{200}$ is the virial mass of the halo. (2) $N_{200}$ is the number of DM particle within the virial radius (4) $r_{200}$. (3) $m_{\rm DM}$ is the mass of a DM particle. (5) $c$ is the halo concentration parameter. (6) $\rho_{\rm s}$ and (7) $r_{\rm s}$ are the scale density and radius of the NFW profile. (8) $\epsilon$ is the gravitational softening length. (9) $\sigma \, m_{\rm \chi}^{-1}$ is the cross section per unit of mass of the SIDM collisions. (10) $t_{\rm col}$ is the collapse time scale, defined in Equation \ref{eqn:collapse_time_scale}. }
    \label{tab:runs_config}    
\end{table*}

We perform dark-matter-only simulations of isolated halos using the multi-physics, massively parallel, moving-mesh magneto-hydrodynamic (MHD) simulation code \textsc{Arepo}, which is publicly available~\cite{Weinberger2020}. Gravity is solved using the Tree-Particle Mesh (Tree-PM) method. Numeric details of the simulations are summarized in Table~\ref{tab:runs_config}. Below we will briefly describe the construction of initial conditions and the modeling of DM self-interactions.

The isolated halos are initialized with $3 \times 10^{7}$ DM particles within $r_{200}$ with distribution of particles extending to $3\times r_{200}$. The density profile within $r_{200}$ follows the NFW profile~\cite{NFW1997}. Beyond $r_{200}$, we use an exponential cut-off following \cite{Springel1999}
\begin{equation}
    \label{eqn:nfw_outer_density}
    \rho\left(r\right) = \frac{\rho_{\rm s}}{c \left(1 + c\right)^2} \left(\frac{r}{r_{200}}\right)^{\epsilon_{\rm decay}} \exp\left(-\frac{r-r_{200}}{r_{\rm decay}}\right),
\end{equation}
where $r_{\rm decay}$ is the decay scale chosen to be $r_{200}$ and $\epsilon_{\rm decay}$ is chosen so that the continuity of the logarithmic slope of the density profile is preserved
\begin{equation}
    \label{eqn:nfw_decay_exponential}
    \epsilon_{\rm decay} = \frac{r_{200}}{r_{\rm decay}} - \frac{1+3c}{1+c}.
\end{equation}

We assume the dynamical equilibrium of the halo and again use the Eddington distribution function (Equation \ref{eqn:eddington_df}) to calculate the phase-space distribution function. To explicitly sample the positions and velocities of particles, we use a combination of analytical formulas and numerical calculations with cubic spline interpolations. The integrated mass $M \left(r\right)$ and negative gravitational potential $\psi \left(r\right)$ are calculated analytically for $r \leq r_{200}$ and numerically for $r > r_{200}$. Following the procedure in \cite{Kazantzidis2004}, we obtain the density profile as a function of negative gravitational potential, $\rho \left(\psi\right)$, from the interpolation of tabulated values of $\rho \left(r\right)$, $M \left(r\right)$, and $\psi \left(r\right)$ from $10^{-5}\,r_{200}$ to $10^3\,r_{200}$. 
We validate our calculation by generating the halo density profile using the acquired Eddington's distribution function. The generated density profile converges almost perfectly with the analytical formulas in equation \ref{eqn:nfw_inner_density} and \ref{eqn:nfw_outer_density}. The relative error is typically orders of magnitude below 10$^{-2}$ for $r > 10^{-4}\,r_{200}$ but is significantly higher for $r < 10^{-4}\,r_{200}$ which can be ignored as the region is sufficiently small.

We sample the positions of particles with the inversion sampling method using the density profile as the distribution function. From the isotropic assumption, we then sample a random direction for each particle and calculate the Cartesian coordinates. For the velocities, we sample a random direction for each particle. In each radial bin, we sample the velocity magnitudes of particles via rejection sampling with the Eddington distribution function at the bin centers. We adopt stretched Gaussian proposal functions, which are fitted to minimize the rejection rate in each bin. 

The initial conditions of a $10^{7.9}\msun$ halo derived through this approach shown in Figure \ref{fig:ics} are tested by running a CDM simulation for $15\,{\rm Gyr}$ and inspecting the evolution of the density and velocity dispersion profiles. We compare the profiles with their analytical formulas. The analytical density profile is described by Equation \ref{eqn:nfw_inner_density} and Equation \ref{eqn:nfw_outer_density}. We obtain the velocity dispersion profile by solving the Jeans equation~\citep{Binney1987}
\begin{equation}
    \label{eqn:velocity_dispersion}
    \frac{1}{\rho\left(r\right)} \frac{\rm d}{{\rm d}r}\left(\rho\left(r\right) \sigma_{\rm r}^2\left(r\right)\right) = \frac{{\rm d}\psi\left(r\right)}{{\rm d}r}.
\end{equation}

Figure \ref{fig:ics} shows the stability of the $10^{7.9}\msun$ halo in CDM. We find that sampling particles to 3\,$r_{200}$ provide sufficient supporting structures for the halo. There are small fluctuations in the density and velocity dispersion profiles at the outer edge of the halo, which takes a few Gyr to evolve into its equilibrium state. Meanwhile, the central density and velocity dispersion also fluctuate, and eventually a small core forms due to two-body relaxation. Nevertheless, the deviation from the analytical profiles created by this core is of the same order of magnitude as the fluctuation due to statistical noises from the limited number of particles inside the region. After 15\,Gyr, our halo becomes totally stable (as shown in Figure \ref{fig:ics}). Its density and velocity dispersion profiles follow closely the analytical profiles except for the small statistical fluctuation. 

We also compared the particle relative velocity distribution of the sampled halo with the analytical result from Section \ref{sec:analytical}. Using the k-d tree algorithm, we find all particle pairs with distances smaller than $0.01\,r_{\rm s}$ in the central region of the halo ($r_{\rm core} = r_{\rm s}$ or $3\,r_{\rm s}$). This distance is chosen empirically to ensure a statistically significant number of particle pairs while keeping the mixing of particles in different shells at a minimum. The relative velocities of the pairs are then calculated and used to evaluate the relative velocity distribution $p_{v_{\rm rel}}^{\rm core}\left(v_{\rm rel}\right)$. This distribution is equivalent to
\begin{equation}
    \label{eqn:core_relative_velocity_df}
    p_{v_{\rm rel}}^{\rm core}\left(v_{\rm rel}\right) = D \int_0^{r_{\rm core}} 4 \pi r^2 {\rm d}r \rho\left(r\right)^2 p_{v_{\rm rel}}^{\rm shell} \left(v_{\rm rel} \mid r\right),
\end{equation}
where $D$ is a normalizing factor. This relative velocity distribution takes into account the shell's densities which influence the number of particle pairs within each shell. We find that throughout the CDM halo evolution, its relative velocity distribution for both choices of core radius ($r_{\rm s}$ and $3\,r_{\rm s}$) follows strictly (with only some minor fluctuations and deviations) the analytical distributions. 

The CDM halo evolved for 15\,Gyr, at which point it settles into dynamic equilibrium, is used as the initial condition for all of our SIDM simulations. DM self-interactions are simulated in a Monte-Carlo fashion as implemented in \cite{Vogelsberger2012,Vogelsberger2014-sidm}. Scatterings of DM particles are assumed to be elastic and isotropic. The implementation ensures explicit conservation of energy and momentum. The time steps of simulations are regulated by the estimated SIDM scattering rates seen by each particle and are short enough to avoid multiple scatterings per time step. We refer readers to \cite{Vogelsberger2012} for detailed descriptions of the numeric implementation. 

We simulate one SIDM halo with the resonant velocity-dependent cross section and two SIDM halos with velocity-independent cross sections. The velocity-independent cross sections are set at the values of $\sigma_{\rm avg}/m$ and $\sigma_\kappa/m$, calculated from Equation \ref{eqn:average_cross_section} and Equation \ref{eqn:conductivity_cross_section}, respectively. The exact values for these effective cross section amplitudes are shown in Table \ref{tab:runs_config}. 

\section{Results}
\label{sec:results}

\begin{figure}
    \centering
        \includegraphics[width=0.49 \textwidth]{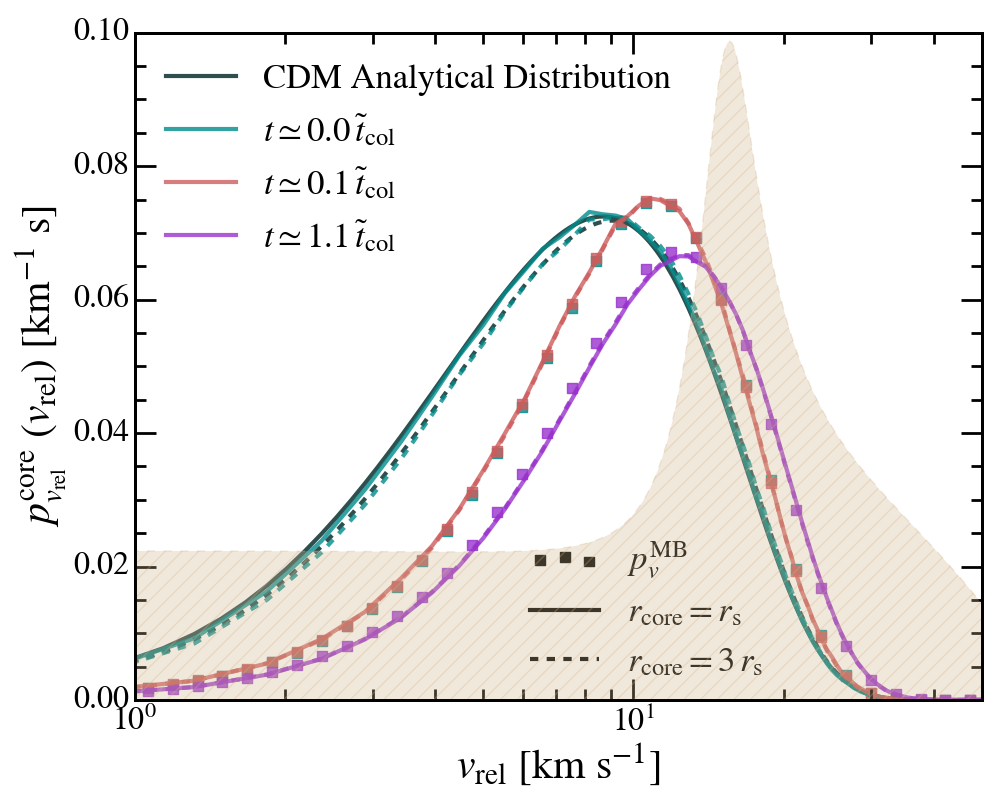}
    \caption{Time evolution of the particle relative velocity distribution inside the halo core $p_{v_{\rm rel}}^{\rm core}\left(v_{\rm rel}\right)$ (Equation \ref{eqn:core_relative_velocity_df}) measured directly from snapshots of the resonant SIDM simulation. Here, distributions from two choices of halo core $r_{\rm core} = r_{\rm s}$ and $r_{\rm core} = 3\,r_{\rm s}$ are shown in solid and dashed lines, respectively. The analytical distributions for CDM halo with NFW profile are also shown in dark blue. As the SIDM halo goes into the core-formation phase, the relative velocity distribution approaches the Maxwell-Boltzmann distribution (shown by squares) calculated from Equation \ref{eqn:Maxwell-Boltzmann_distribution} with the central radial velocity dispersion $\sigma^2_{r,{\rm c}}$ as $v_0^2$. The resonant model cross section profile is shown in yellow and scaled logarithmically in both axices.}
    \label{fig:relative_velocity_evolution}
\end{figure}

\begin{figure}[!h]
    \centering
    \includegraphics[width=1\linewidth]{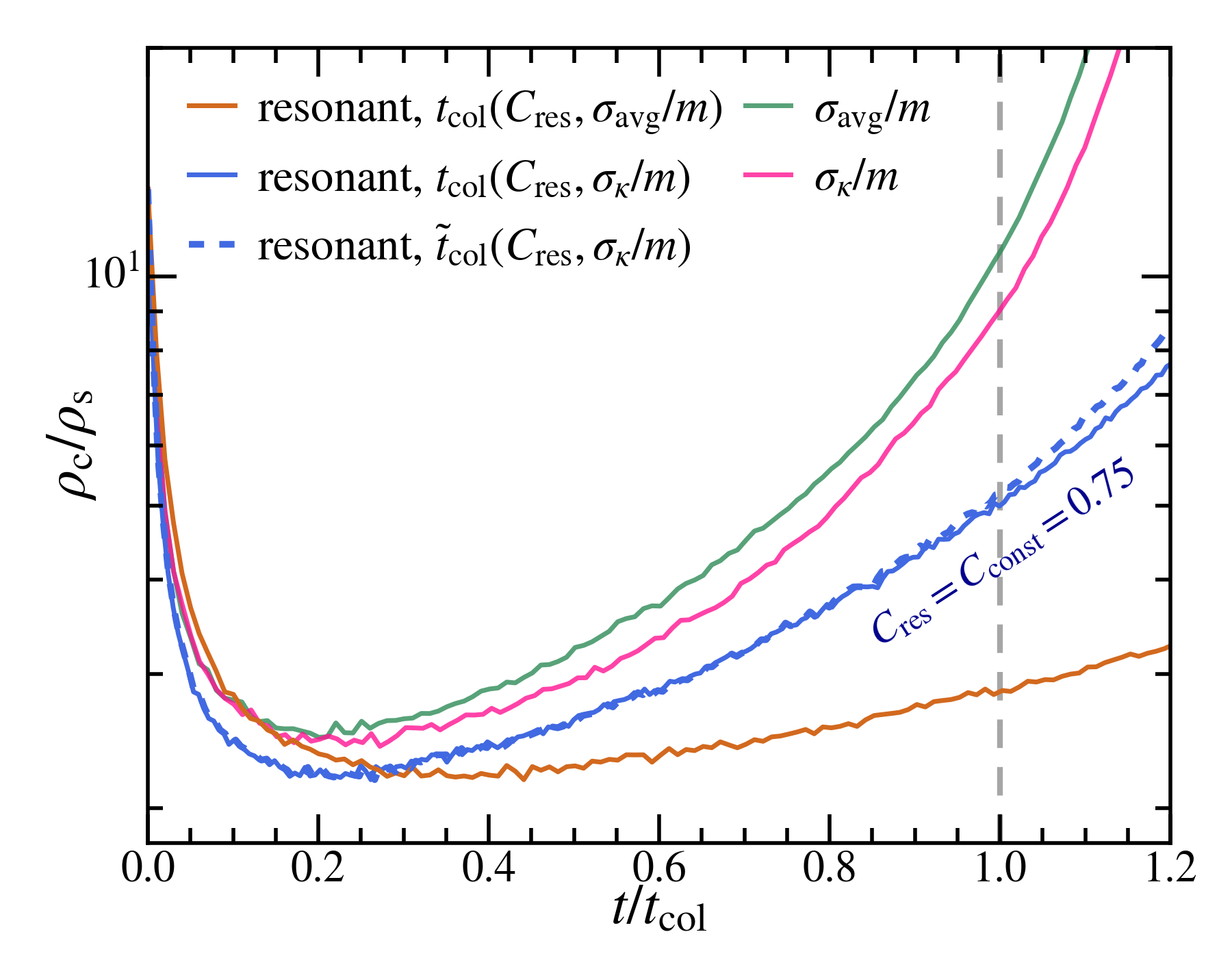}
    \includegraphics[width=1\linewidth]{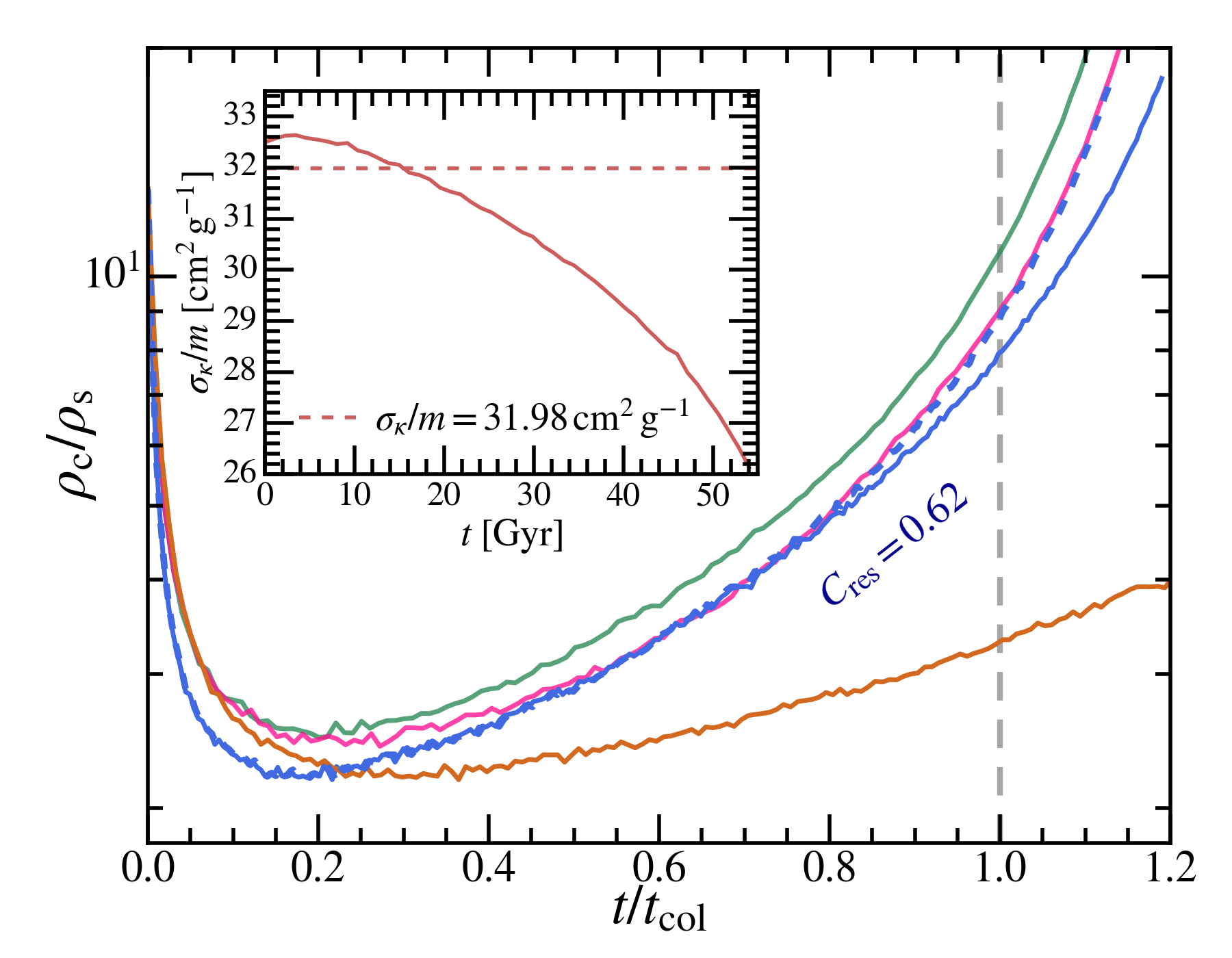}
    \caption{Time evolution of DM central (at $r\leq 0.1\, r_{\rm s}$) densities in the VICS and the RCS SIDM models. The times are normalized by the collapse time $t_{\rm col}$ (as defined in Equation~\ref{eqn:collapse_time_scale}) of the corresponding model. For the halo in the RCS model, we also normalize the collapse time with an adaptive timescale $\tilde{t}_{\rm col}$ described in Equation \ref{eqn:adaptive_time_scale} that take into account of the change in the relative velocity distribution and, consequently, the conductivity cross section $\sigma_\kappa/m$ over time. The top panel shows the results when the same prefactor of conduction coefficient $C=0.75$ is assumed for both VICS and RCS models. The bottom panel shows the results when the prefactor for the RCS model is changed to $C_{\rm res}\sim 0.6$. In general, halos in SIDM undergo a first phase of core formation until thermal equilibrium is reached by DM self-interactions. The collapse time based on $\sigma_{\rm avg}/m$ is better in describing the core formation phase in the RCS model. On the contrary, the later gravothermal collapse phase is better described by $\sigma_{\kappa}/m$ despite about $20\%$ delay in collapse time, which we attempt to capture by adjusting $C_{\rm res}$ as indicated. This adjustment also better matches the normalized central density evolution described by $\sigma_{\rm avg}/m$ to the core formation phase.}
    \label{fig:central-density-evol}
\end{figure}

\begin{figure}[!h]
    \centering
    \includegraphics[width=0.98\linewidth, trim={0.25cm 0 0 0}]{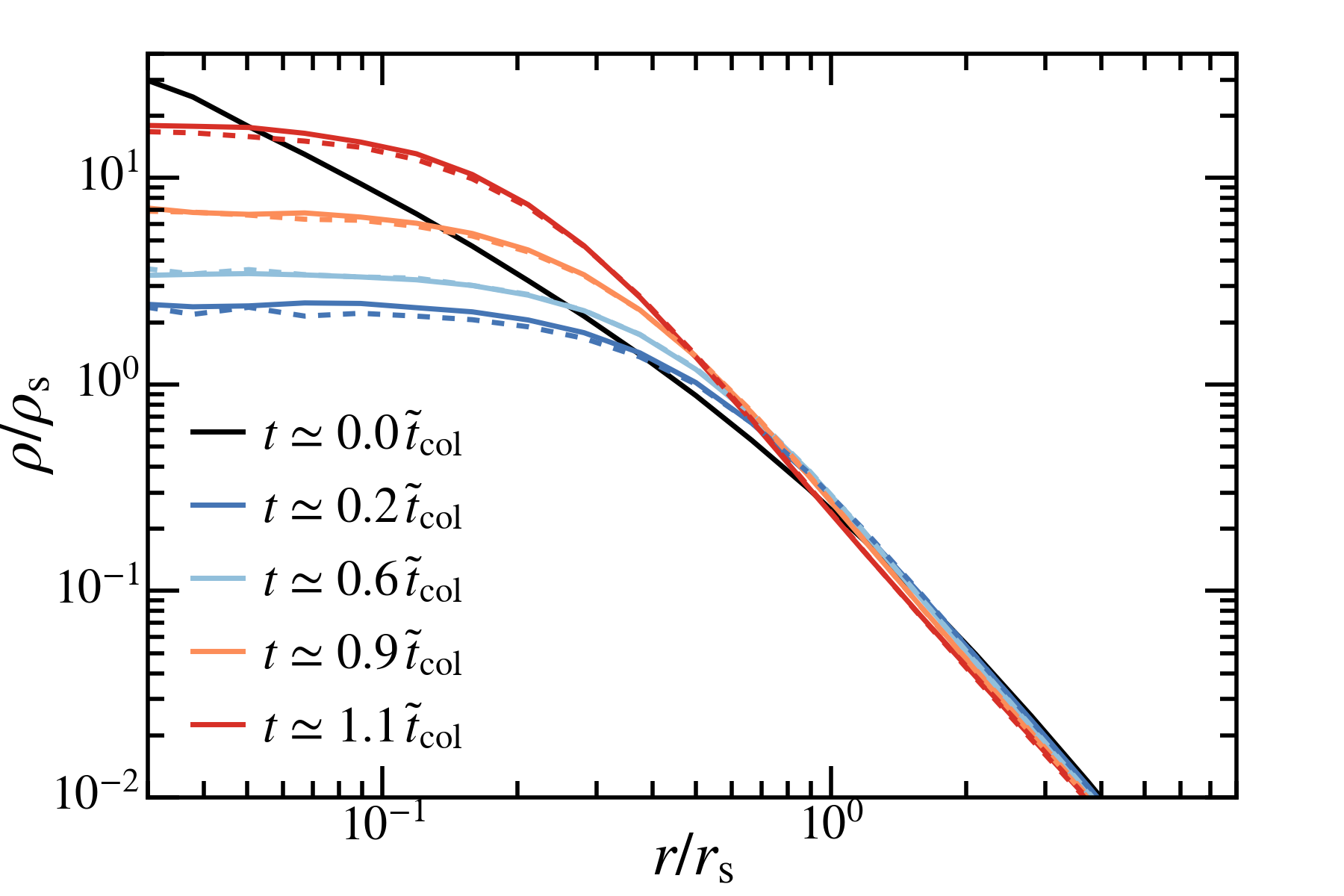}
    \includegraphics[width=1\linewidth]{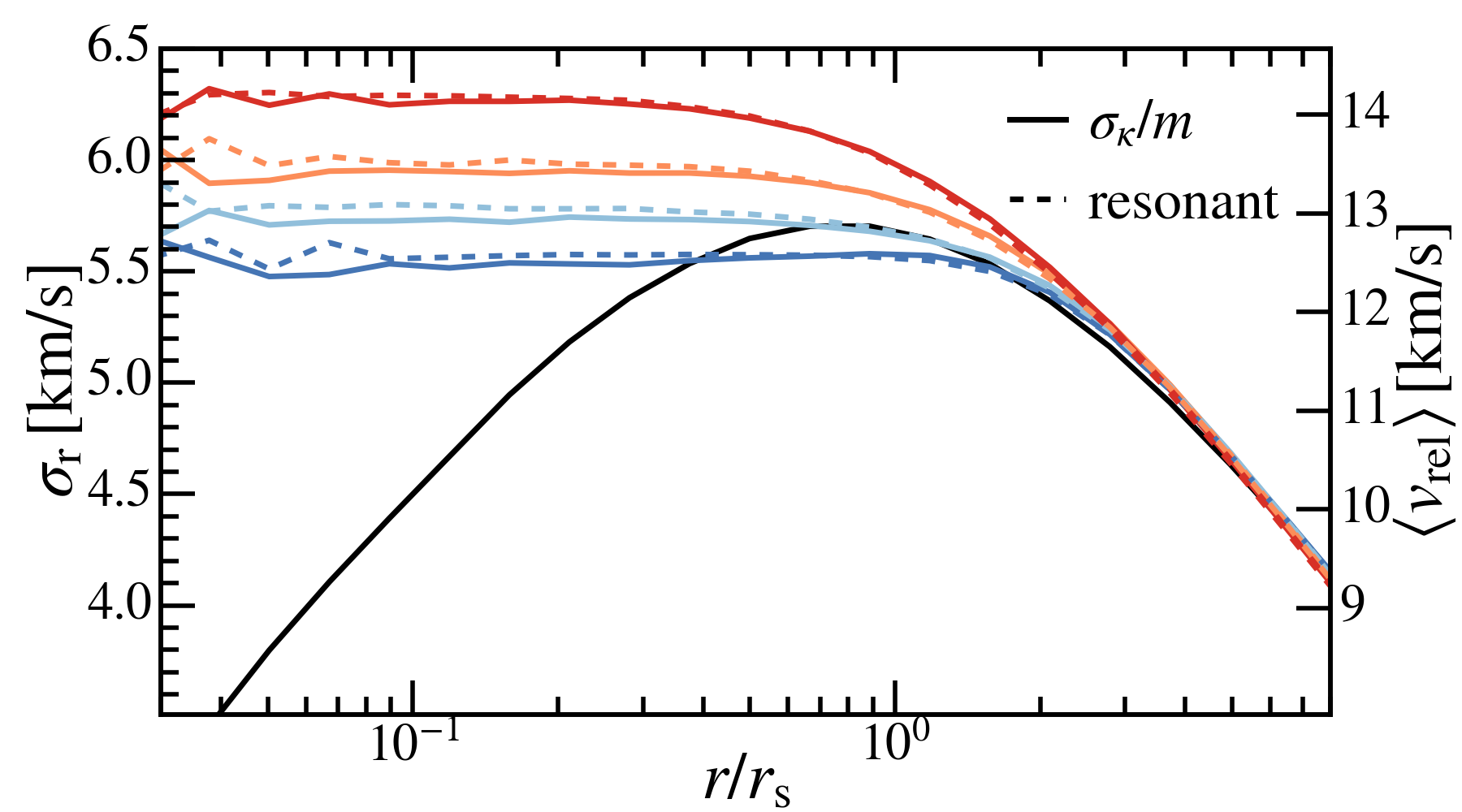}
    \includegraphics[width=1\linewidth]{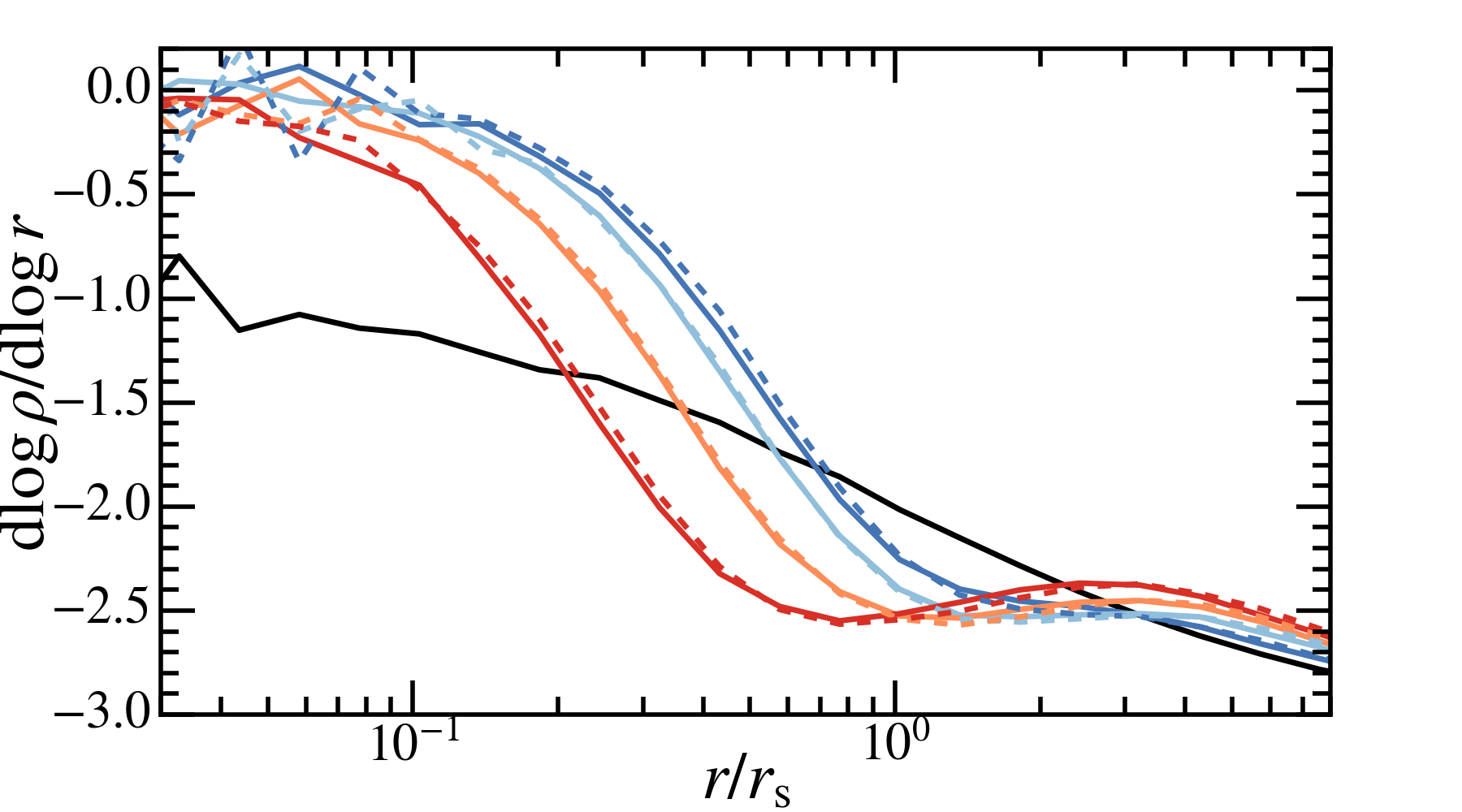}
    \caption{Time evolution of DM density profiles (top), velocity dispersion profiles (middle), and logarithmic profile slopes (bottom) of the simulated halo in the VICS and RCS SIDM models. The mean relative velocity $\langle v_{\rm rel}\rangle \equiv \sigma_{\rm r} \times 4/\sqrt{\pi}$, assuming a Maxwell-Boltzmann distribution, is shown in the bottom panel as a reference. The qualitative evolution of halo structure in the VICS and RCS models is similar. The rapid core-formation phase ($t\lesssim 0.2\,t_{\rm col}$) is followed by the gravothermal collapse phase, where the halo centers remain isothermal and the density profiles are self-similar. The velocity dispersion profile of the halo increases mildly through the collapse phase. Quantitatively, such a small ($\lesssim 1\kms$) increase in $\langle v_{\rm rel} \rangle$ does not change the effective cross-section and conduction efficiency substantially. Therefore, a fixed $t_{\rm col}$ or equivalently $C_{\rm res}$ can approximate the entire collapse phase.}
    \label{fig:density-profile}
\end{figure}

In this section, we inspect the evolution of the halo in the velocity-independent cross section (VICS) and resonant cross section (RCS) SIDM models. 

Figure \ref{fig:relative_velocity_evolution} shows the relative velocity distributions $p_{v_{\rm rel}}^{\rm core}\left(v_{\rm rel}\right)$ of the halo in the RCS model at 0, 0.1, and 1.1 $\tilde{t}_{\rm col}$ with $\tilde{t}_{\rm col}$ as the adaptive time scale defined below in Equation \ref{eqn:adaptive_time_scale}. As mentioned in Section \ref{sec:simulation}, this distribution is read directly from the simulation snapshots and takes into account the weighting factor of the local particle density. We observe that the relative velocity distribution rapidly transforms into Maxwell-Boltzmann distributions (Equation \ref{eqn:Maxwell-Boltzmann_distribution}) with the central radial velocity dispersion $\sigma_{r,{\rm c}}$ as as the characteristic velocity $v_0$ in the core formation phase and remain so during the core collapse. These Maxwell-Boltzmann distributions are shown by squares in Figure \ref{fig:relative_velocity_evolution}.

In Figure \ref{fig:central-density-evol}, we show the evolution of halo central density versus time in the RCS and two VICS models. The time is normalized by the collapse timescale $t_{\rm{col}}$, as defined in Equation~\ref{eqn:collapse_time_scale}. The central density of each halo is calculated as the average DM density within $r_{\rm c} = 0.1\,r_{\rm s}$~\footnote{The value of $r_{\rm c}$ is chosen to reduce statistical uncertainties while keeping the region small enough to represent the halo central density.}. The top panel shows the evolution of halos with collapse timescales calculated using two different definitions of the effective cross section ($\sigma_{\rm avg}/m$ and $\sigma_{\kappa}/m$) and $C = 0.75$. The green and pink curves represent halos evolving with VICS of amplitudes $\sigma_{\rm{avg}}/m$ and $\sigma_{\kappa}/m$, respectively. The orange (blue) curve shows the evolution of the halo in the RCS model when using $\sigma_{\rm{avg}}/m$ ($\sigma_{\rm{\kappa}}/m$) to compute $t_{\rm{col}}$. 

In addition, we define an adaptive time scale $\tilde{t}_{\rm{col}}$ to account for the change of $\sigma_{\rm{\kappa}}/m$ over time due to the shift of $p_{v_{\rm rel}}^{\rm core}\left(v_{\rm rel}\right)$ as seen in Figure \ref{fig:relative_velocity_evolution}. The phase of collapse, $t/\tilde{t}_{\rm{col}}$, is defined following
\begin{equation}
    \label{eqn:adaptive_time_scale}
    {\rm d} \left(\frac{t}{\tilde{t}_{\rm{col}}}\right) = \frac{{\rm d}t}{t_{\rm col} \left(\sigma_\kappa\left(t\right)/m\right)}.
\end{equation}
where $t$ is the physical time, and $\sigma_\kappa\left(t\right)/m$ is calculated from Equation \ref{eqn:conductivity_cross_section} using $p_v^{\rm MB}\left(v\right) = p_{v_{\rm rel}}^{\rm core}\left(v_{\rm rel}\right)$ as measured according to the procedural detailed in Section \ref{sec:simulation}. The insert in the bottom panel of Figure \ref{fig:central-density-evol} shows the evolution of the conductivity effective cross section $\sigma_\kappa\left(t\right)/m$ over physical time. Generally, $\sigma_\kappa\left(t\right)/m$ evolve proportional to $\sim t^{-2}$, decreasing from $\simeq 33.5\cpm$ at $t = 0$\,Gyr to $\simeq 26\cpm$ at the time the central density reach its initial value $t \sim 55$\,Gyr. The dashed blue line in Figure~\ref{fig:central-density-evol} shows the evolution of the central density in the halo in the RCS model rescaled with $\tilde{t}_{\rm col}$. 

As expected, the halos evolving with velocity-independent cross sections exhibit nearly universal evolution when time is scaled by $t_{\rm{col}}$, with a small deviation in the core-collapse phase likely due to numerical effects \citep[e.g.][]{Mace+24, Palubski2024, Shen2024}. For the resonant model, we find that computing $t_{\rm{col}}$ with $\sigma_{\rm avg}/m$ (orange curve) provides a better description of the initial core-formation phase until $t \sim 0.2 \ t_{\rm{col}}$, but the agreement breaks down for $t \gtrsim 0.2 \ t_{\rm{col}}$. In the later stages of evolution, after the core-formation phase, calculating $t_{\rm{col}}$ using $\sigma_{\kappa}/m$ (blue curve) more closely tracks the evolution of the velocity-independent models. However, the halo evolving with a resonant cross section still takes $\sim 20\%$ longer in rescaled time units $t / t_{\rm{col}}$ to reach its initial central density during the collapse phase than the halos evolving with VICS. This is also the case for the evolution in the RCS model scaled by the adaptive time unit $t / \tilde{t}_{\rm{col}}$. 

In the bottom panel of Figure \ref{fig:central-density-evol} we show the blue and orange curves for the resonant model after calculating $t_{\rm{col}}$ with $C_{\rm res} = 0.62$, which one could also interpret a modification of $\sigma_{\rm{eff}}$ at the level of $\sim 20 \%$ to match the evolution in the core collapse phase. Using $C_{\rm res} = 0.62$ with the adaptive time scale $\tilde{t}_{\rm{col}}$ as shown by the dashed blue line in the bottom panel of Figure \ref{fig:central-density-evol} maps the central density evolution in the RCS model exactly along the rescaled collapse track of the halo with VICS of amplitude $\sigma_\kappa/m$ for $t \gtrsim 0.4\,t_{\rm{col}}$. In addition, the choice of $C_{\rm res} = 0.62$ also better matches the central density evolution of the halo in the RCS model when rescaled with $t_{\rm{col}}\left(\sigma_{\rm avg}/m\right)$ to the universal track in the core formation phase.

However, even after introducing this ad-hoc rescaling of the collapse timescale, as well as utilizing different time scales to describe different phases, the behavior during the late stage of core formation and early stage of core collapse ($0.2\,t_{\rm col} \lesssim t \lesssim 0.4\,t_{\rm col}$) still differ from those of the halos we evolve with velocity-independent cross sections. The minimum core density of the halo in the RCS model is smaller than that in the VICS model by a factor of $\simeq 10 \%$. We interpret these deviations as evidence of a violation of universality at the level of $\sim 10-20 \%$ for halos evolving with resonant cross sections.

The evolution of the density and velocity dispersion profiles are shown in Figure~\ref{fig:density-profile}. We show the DM density and relative velocities at several snapshots in one velocity-independent model (the halo with a velocity-independent cross section $\sigma_{\kappa}/m$) and the resonant model. Based on Figure \ref{fig:central-density-evol}, we use $\tilde{t}_{\rm col}$ and $C_{\rm res}= 0.62$ to define the collapse timescale for the resonant model to match the evolution during the collapse phase seen in the velocity-independent models. 

\section{Discussion and conclusions}
\label{sec:conclusions}
We have performed idealized N-body simulations of DM halos to examine halo evolution in a SIDM model with a resonant cross section. We compare the evolution of the halo with a resonant cross section with the evolution of halos evolving with velocity-independent cross sections, and examine the change in their internal structure over time using two definitions of an effective cross section and core collapse timescale. Our main findings are summarized as follows: 
\begin{enumerate}
    \item The resonant model exhibits deviation from universality in the evolution of its density profile from halos evolving with velocity-independent cross sections. The halo evolving with a resonant cross section achieves a minimum core density $10 \%$ lower than the halos with velocity-independent scattering cross sections, and reaches its initial density $20 \%$ later when time is scaled by the thermal conduction cross section $\sigma_{\kappa}\sim \langle \sigma v^5\rangle / \langle v^5 \rangle $. 
    \item  In the early thermalization and core-formation phase in the resonant model, the halo evolution is better described by the averaged cross-section ($\sigma_{\rm avg}/m$) ``seen'' by particles having self-interactions. After core formation, the late gravothermal collapse phase in the resonant model is better described by $\sigma_{\kappa}/m$, although the time evolution resulting from this definition of the effective cross section still deviates significantly from the velocity-independent models.
    \item For our specific resonant SIDM model and halo mass choice, using $C_{\rm res} = 0.62$ matches the central density evolution when rescaled with $t_{\rm col}\left(C_{\rm res},\sigma_{\rm avg}\right)$ exactly with the self-similar track in the core formation phase. Additionally, when rescaled with the adaptive time scale $\tilde{t}_{\rm col}\left(C_{\rm res},\sigma_\kappa\right)$, the evolution of the central density follows precisely along that of the halo evolved with a velocity-independent cross section of amplitude $\sigma_\kappa$ during most of the core collapse.
\end{enumerate}
The utility of a universal evolution of SIDM halos lies in its predictive power; given the physical properties of a CDM halo ($\rho_s$ and $r_s$) and the form of the SIDM cross section, the universal treatment of halo evolution based on the collapse timescale enables the prediction of minimum core density and the structure of the halo during the collapse phase \cite{Yang+2022}. While we observe a deviation from universality, it is only at the level of $10 - 20\%$. Other physical processes can impact SIDM halo density profiles to a similar or even larger degree, particularly tidal stripping and tidal heating effects relevant for galactic subhalos \citep[e.g.][]{Nishikawa+20,Zeng+22,Slone+23}.

A modified calculation of the core collapse timescale based on a time-dependent definition of the thermal conduction cross section $\sigma_{\kappa}$ provides an improved description of the collapse phase for the halo evolving with a resonant cross section. This suggests that a modified calculation of the thermal conduction cross section could restore universality for resonant cross sections, provided this modified treatment reverts to the definition of $\sigma_{\kappa}$ for velocity-independent cross sections. Additionally, provided analyses are robust to systematic uncertainties at the level of $\sim 20\%$ in the collapse timescale, studies involving dwarf galaxy observations and gravitational lensing could distinguish these classes of models from collisionless dark matter, improving on the existing constraints from lensing on RCS models presented by \cite{Gilman+23}. Further exploration of halo evolution with resonant SIDM cross sections with different resonant structures across a wider range of halo masses will be presented in a forthcoming publication.

\begin{acknowledgments}
We thank Annika Peter, Antti Asikainen, and Yi-Ming Zhong for useful discussions. We thank the anonymous referee for a constructive report.
DG acknowledges support for this work provided by the Brinson Foundation through a Brinson Prize Fellowship grant. 
MV acknowledges support through NASA ATP 19-ATP19-0019, 19-ATP19-0020, 19-ATP19-0167, and NSF grants AST-1814053, AST-1814259, AST-1909831, AST-2007355 and AST-2107724.
\end{acknowledgments}

\nocite{*}

\bibliography{main-apsformat}

\end{document}